\documentclass[sigconf]{acmart}

\usepackage{booktabs} 
\usepackage{algorithm}
\usepackage{algpseudocode}
\usepackage{multirow}

\DeclareMathOperator*{\argmin}{argmin}

\setcopyright{rightsretained}

\acmConference[WSDM'18]{ACM International Conference on Web Search and Data Mining}{Feb 2018}{Los Angeles, California, USA} 
\acmYear{2018}
\copyrightyear{2018}
\begin{document}
\title{Fine-Grained Retrieval of Sports Plays using Tree-Based Alignment of Trajectories}

\author{Long Sha}
\affiliation{%
  \institution{Qld Uni of Tech/STATS}
  \city{Brisbane} 
  \country{Australia} 
}
\email{long.sha@connect.qut.edu.au }

\author{Patrick Lucey}
\affiliation{%
  \institution{STATS}
  \city{Chicago} 
  \country{USA} 
}
\email{plucey@stats.com}

\author{Stephan Zheng}
\affiliation{
  \institution{Caltech}
  \city{Pasadena}
  \country{USA}}
\email{stzheng@caltech.edu}

\author{Taehwan Kim}
\affiliation{
  \institution{Caltech}
  \city{Pasadena}
  \country{USA}}
\email{taehwan@caltech.edu}

\author{Yisong Yue}
\affiliation{%
  \institution{Caltech}
  \city{Pasadena} 
  \country{USA}}
\email{yyue@caltech.edu}

\author{Sridha Sridharan}
\affiliation{
  \institution{Qld Uni of Tech}
  \city{Brisbane}
  \country{Australia}}
\email{s.sridharan@qut.edu.au}

\begin{abstract}
We propose a novel method for effective retrieval of multi-agent spatiotemporal tracking data. Retrieval of spatiotemporal tracking data offers several unique challenges compared to conventional text-based retrieval settings.  Most notably, the data is fine-grained meaning that the specific location of agents is important in describing behavior. Additionally, the data often contains tracks of multiple agents (e.g., multiple players in a sports game), which generally leads to a permutational alignment problem when performing relevance estimation. Due to the frequent position swap of agents, it is difficult to maintain the correspondence of agents, and such issues make the pairwise comparison problematic for multi-agent spatiotemporal data. 
To address this issue, we propose a tree-based method to estimate the relevance between multi-agent spatiotemporal tracks. It uses a hierarchical structure to perform multi-agent data alignment and partitioning in a coarse-to-fine fashion. We validate our approach via user studies with domain experts. Our results show that our method boosts performance in retrieving similar sports plays -- especially in interactive situations where the user selects a subset of trajectories compared to current state-of-the-art methods.  
\end{abstract}

%
%


\maketitle
%
\section{Introduction}

Research into ``exemplar-based'' and ``sketched-based'' approaches to image retrieval has recently surged~\cite{bui2015scalable,yu2016sketch,zhang2016sketchnet}. The recent popularity of these \emph{fine-grained retrieval} methods is due to the inadequacies of current  ``text-based'' or ``key-word'' query-based methods. As a \emph{picture tells a thousand words}, using examples or sketches which capture the fine-grained attributes that the user is interested in has shown to be superior to text-based searches.

\begin{figure}[t]
\centering
\includegraphics[width=\linewidth]{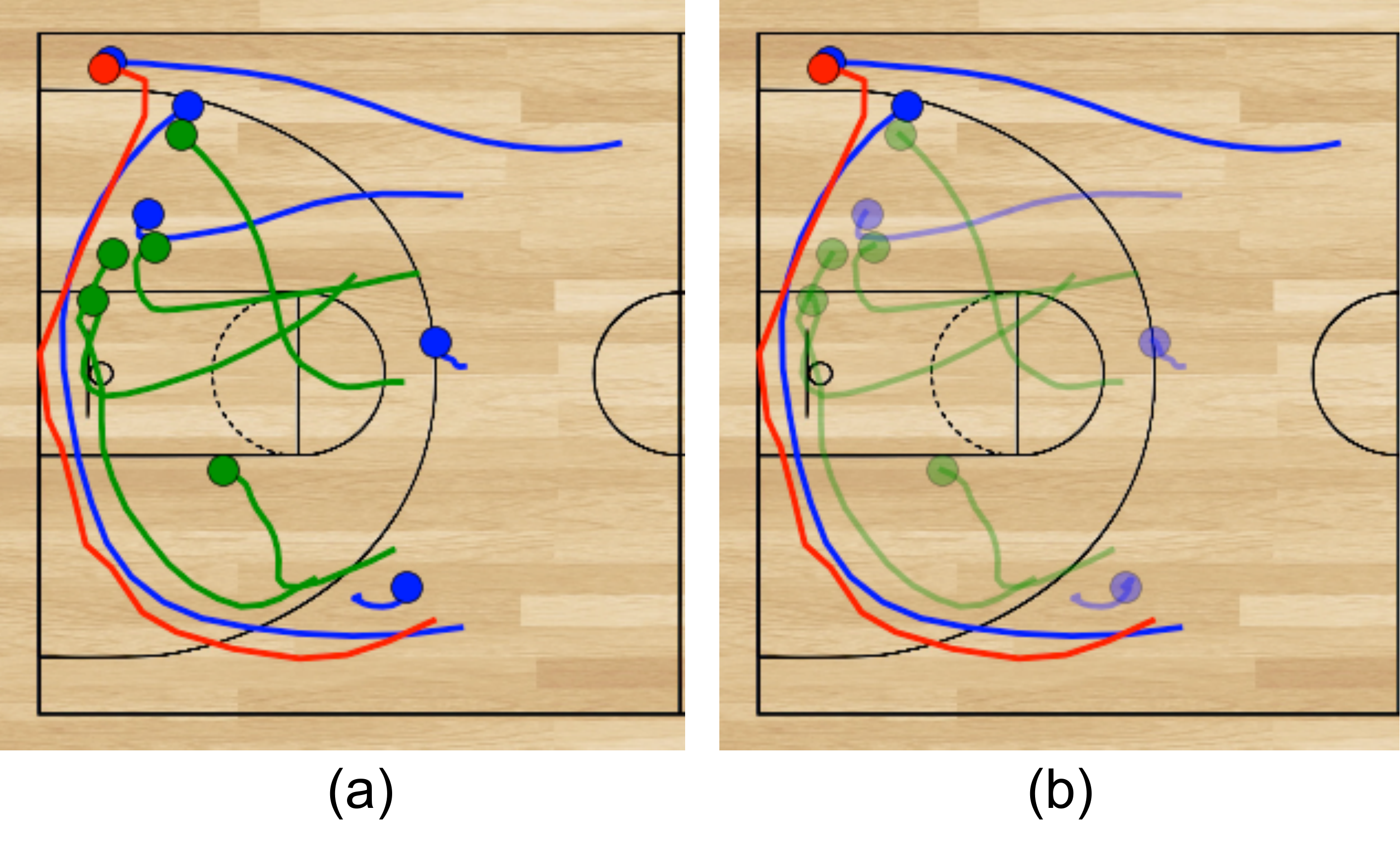}
\caption{In this paper we focus on retrieving fine-grained multi-agent data in basketball using a tree-based alignment method. We focus on two types of retrieval tasks: (a) Given a four-second example play (blue is offensive team, green is defense and red is the ball -- the small circle on each trajectory shows the end point) we use that as the input query and retrieve all plays that look similar to that query; and (b) Given the same play, the user selects only the trajectories of interest and the retrieval is based on that chosen subset.}
\label{fig:query_example}
\end{figure}

We study the setting of fine-grained retrieval of multi-agent spatiotemporal data such as sports plays. A depiction of the ``exemplar search'' problem is shown in Figure~\ref{fig:query_example}(a). Given an input play of a specific length (say 4 seconds), the input query is then compared to the entire database of 4 second plays and a ranked list of most similar plays are then retrieved. Additionally, the user selects only the players that they are interested in (Figure~\ref{fig:query_example}(b)). Based on the subset of players, the system then retrieves a ranked list of similar plays based only on the selected trajectories. Previous work showed that users much preferred the exemplar and sketched-based method over conventional keyword-based retrieval system \cite{sha2016chalkboarding}. Most crucially, the retrieval system allowed users to retrieve fine-grained plays \emph{in a matter of seconds, instead of days} which can often be the case in practice in sports domains.  



One major challenge in relevance estimation for multi-agent trajectories is that of alignment.
Although low-dimensional compared to using an image-based representation, the inherent problem of using the raw multi-agent data is that of the misalignment due to the constant swapping of player positions over the course of a play (i.e., permutation problem). One approach to circumventing this issue is to utilize a preprocessing step which pre-aligned the multi-agent data to a template was utilized which allowed for quick local trajectory comparison. 

\begin{figure*}[!t]
\centering
\includegraphics[width=0.9\linewidth]{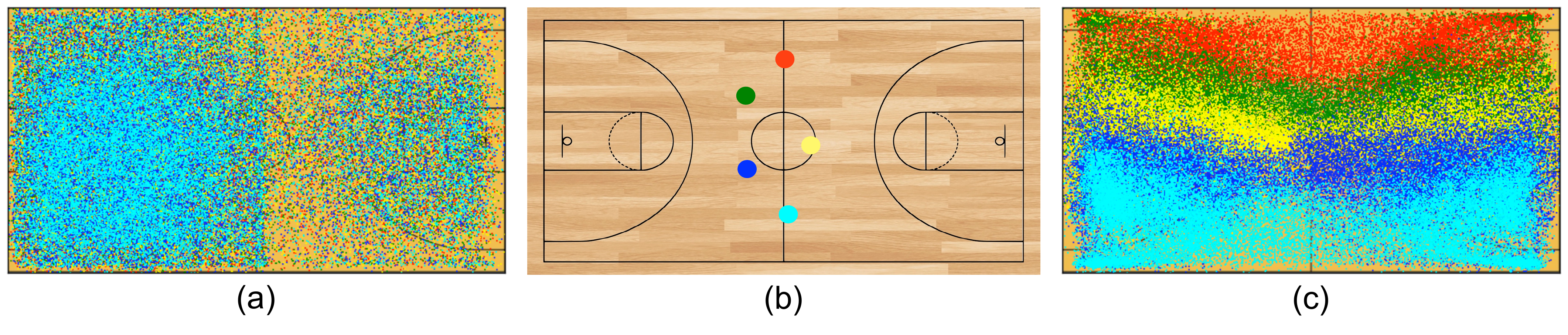}
\caption{(a) Given an initial ordering with a color corresponding to each player in a team, we show that a player's position across a quarter of a game is quite random. (b) But if we align or permute the ordering at the frame-level to this template, we can (c) discover the hidden structure of the team. In this plot we show the alignment to a single template, but we will show later than using a tree of templates is more effective.}
\label{fig:alignment}
\end{figure*}

In this paper, we propose an improved multi-agent data alignment method which gives improved fine-grained retrieval performance. Our proposed method uses a hierarchical structure to perform multi-agent data alignment and partitioning in a coarse-to-fine fashion. Our approach can be easily integrated into existing spatiotemporal retrieval pipelines.  We validate our approach using over a wide range of retrieval tasks.  Our user study results demonstrate significant benefits of our method over previous relevance estimation methods for sports play retrieval.

The rest of the paper is as follows. In Section 2 we describe the importance of aligning multi-agent data and why using the raw data is preferred. In Section 3, we explain our proposed tree-based multi-data alignment method, and Section 4 describes how this is implemented within a retrieval framework. Section 5 shows our results, and Section 6 gives the relevant related work. We conclude with a summary and discuss future work.\footnote{A demo video of our work can be viewed in \url{https://www.dropbox.com/s/dbhutkek9tz8rpk/WSDM_video.mp4?dl=0}}

\section{Measuring Similarity via the Alignment of Raw Multi-Agent Data}

For effective retrieval to take place, we need an accurate and efficient similarity measure between multi-agent inputs. As shown in  Figure~\ref{fig:alignment}(a), if we look at the raw positional data of a single team in basketball (i.e., 5 players) across a quarter of a match, given an initial ordering we can see that this ordering of the positional data contains little team structure as players tend to constantly switch positions. To counter this issue, we could exhaustively compare the pairwise distance between each player in the input query to a candidate play in the database which is manageable ($5!=120$). But if we include the other team, the exhaustive approach starts to get prohibitive\footnote{In basketball, there are 5 players per team. To compare the offensive team trajectories (i.e., team with the ball), there are $5!=120$ permutations or comparisons required. To include the defensive team, we square this - $(5!)^2=14,400$. We then add the ball trajectory comparison which yields 14,401 comparisons. For other team sports such as soccer which have a higher number of players (i.e., 10 on-field players), the number of permutations are higher than the number of atoms in the universe - $(10!)^2$}.


A solution to bypass the alignment issue is to use hand-crafted features~\cite{li2010group,stracuzzi2011application,wang2004automatic,wei2014forecasting}. Alternatively, a more intuitive approach would be to compare plays as images as it is a visual data source and it has been employed previously for multi-agent data~\cite{yue2014learning,zheng2016generating,miller2014factorized}. 

To employ an image-based approach we can do the following. Say we have a play across a window of time, like four-seconds, and the player and ball information is captured at 25 frames-per-second, we first quantize a court into a series of $1\times1$ foot cells (each one being a pixels).  For a $94\times50$ feet basketball court, this would result in a $94\times50\times3$ RGB image ($D=14,100$), with each channel being assigned to each team (offensive=blue, defense=green and ball=red). In addition to being high-dimensional, it is also lossy, meaning if we wanted to reconstruct the original signal this would be problematic as we have thrown away the temporal structure (i.e., we would not know which location each player was at in each frame). To maintain the original temporal structure, we could add another channel which would result is an extremely sparse and even higher-dimensional input signal -- $D\approx10^{6}$. For a retrieval system, this is highly undesirable as this would require us to store this high-dimensional data in addition to the raw data. 

However, this approach is unnecessary when one considers that the original input data to create the image is already super compact and can be described via a dense matrix of spatial positions. For example, given the $x,y$ location of all 10 players and the $x,y,z$ of the ball, we can represents each frame as a $M=23$ dimensional vector. Across $F$ frames, the multi-agent behavior can be represented as a $M\times F$ matrix -- which in this case would be a dimensionality of $23\times100=2300$.  To utilize the raw data, a solution is to align the raw positional data to a \emph{role template}. This method was proposed by Lucey et al.~\cite{lucey2013representing}, which dynamically assigns an unique role to each agent in each frame according to a single template.

Figure~\ref{fig:alignment}(b) shows the formation template of role-based alignment and (c) shows the \emph{aligned} player positions once this method has been applied. The last plot clearly shows that some type of team structure is obtained. In terms of retrieval, this means that once the permutation matrix has been applied - only a single comparison between trajectories needs to be made. Additionally, only the permutation matrix needs to be stored and not a high-dimensional representation like an image. 


Even though effective, as can been seen in Figure~\ref{fig:alignment}(c), the role-based method is suboptimal as it only picks up the coarse structure of a team. In this figure, we see the thin strips across the court which does not coincide with any meaningful interpretation of the game of basketball. A more meaningful representation would pick up the typical defensive and offensive structures. In the next section, we show how we can do this which yields better retrieval performance. 



\section{Tree-Based Alignment}

We now describe our main technical contribution.
For effective retrieval using raw multi-agent data, accurate alignment is required. At a high-level, this means that we want to find the ordering of players in the input query to the candidate play which minimizes the difference between the two. Technically, this refers to finding the permutation matrix $\mathbf{P}()$ that minimizes the $L_{2}$ distance between all the agents in one team 
\begin{equation}
\argmin_{\mathbf{P}} ||\mathbf{P}(\mathbf{X}_{\mathrm{query}})- \mathbf{X}_{\mathrm{play}}||_2
\label{eq:query_obj}
\end{equation}
where $\mathbf{X}_{\mathrm{query}}$ is the matrix of the spatial positions of agents in the input query in one team with an initial agent order (i.e., that order is fixed across that window of time), $\mathbf{X}_{\mathrm{play}}$ is the matrix of spatial positions of agents in order according to a pre-aligned template for a candidate play in the database, and permutation matrix $\mathbf{P}()$ indicates the correspondence of agents between $\mathbf{X}_{\mathrm{query}}$ and $\mathbf{X}_{\mathrm{play}}$.


In terms of pre-aligning the data in the database, a similar approach is used but instead of just applying the permutation matrix to the input query, we find the permutation matrix between every input play in the database and the gold-standard template, $\mathbf{X}_{\mathrm{template}}$. To determine the gold-standard template, this can be either hand-crafted by a domain~\cite{lucey2013representing} or learnt in a data-driven method using the EM algorithm~\cite{bialkowski2014large}.


\begin{figure}[t]
\centering
\includegraphics[width=0.8\linewidth]{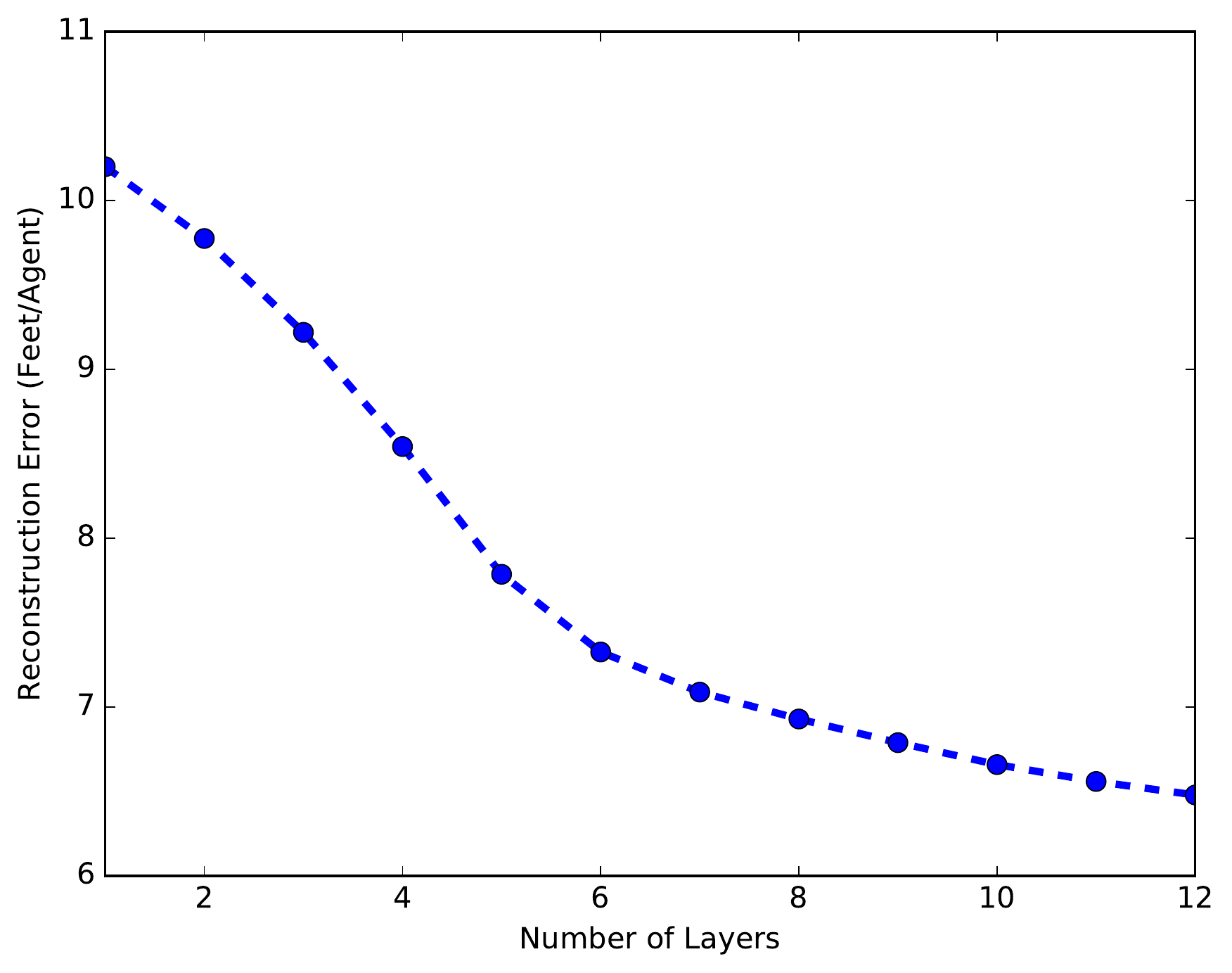}
\caption{The reconstruction cost in Equation \ref{eq:cluster_obj} at each layer. We set the minimum depth to 6 since it shows as the elbow point in this plot.}
\label{fig:depth_test}
\end{figure}

The above approach has yielded reasonable performance but it assumes the observed behavior is linear (i.e., single state). In complex scenarios like those that exist in a team sport like basketball, it is a more reasonable assumption that the behaviors are non-linear which consists of many states. As such, it would make intuitive sense that a superior approach is to learn a separate template for each of these states.  

As the various game-states are not explicit, we can use hierarchical clustering to discover these latent states to provide better alignment. However, this presents a ``chicken-or-the-egg'' problem, as we can not cluster the multi-agent data without it being aligned first. As such, we can apply a coarse-to-fine approach where we first align the data to a coarse template, and then using this initial alignment we can partition the data into finer states which provide templates which allow us to find a better alignment.  

A feasible method to do this would be to use a tree approach, which iteratively executes two steps: alignment and data partitioning. The ultimate goal is to find a set of states/templates that can reasonably reconstruct the complex multi-agent behaviors, and align each data point with the corresponding states/template.

 \begin{figure*}[t]
\centering
\includegraphics[width=\linewidth]{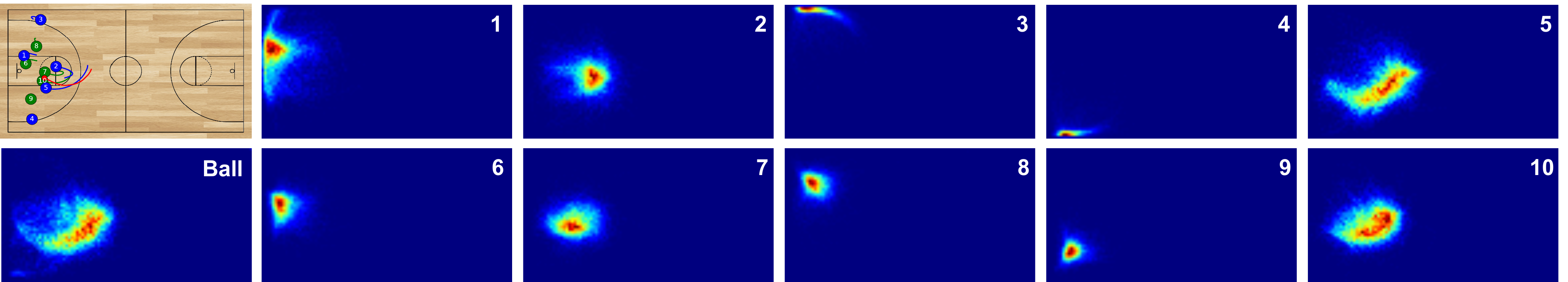}
\caption{An leaf node example and the distribution heat map of each agent. The top left image shows the centroid of this leaf node while others shows the agent distribution heat map within this node.}
\label{fig:leaf_heatmap}
\end{figure*}

\begin{algorithm}[t]
\caption{Template Learning Algorithm}\label{alg:template_learn}
\begin{algorithmic}[1]
\Procedure{TemplateLearn($\mathcal{X}$)}{}
\State Initialize the template with a randomly selected example $\mathbf{X}_{\mathrm{template}}=\hat{\mathbf{X}}$ 
\While{$\Delta \mathbf{X}_{\mathrm{template}}\ <\ threshold$ or $iteration\  <\  max$}
\For{Each sample $\mathbf{X}_i$ in dataset $\mathcal{X}$}
\State Calculate $\mathrm{cost}(m,n) =  ||\mathbf{X}_{\mathrm{template}}(m) - \mathbf{X}_i(n)||_2$ for each agent-agent pair
\State Compute $\mathbf{P}()$ using Hungarian algorithm.
\State Align the example $\mathbf{X}_{i}=\mathbf{P}( \mathbf{X}_i)$
\EndFor
\State Update the template $\mathbf{X}_{\mathrm{tempalte}}$ by averaging aligned $\mathcal{X}$
\State Compute the difference $\Delta \mathbf{X}_{\mathrm{template}}$
\EndWhile \\
\Return {$\mathbf{X}_{\mathrm{template}}$}
\EndProcedure
\end{algorithmic}
\end{algorithm}

\subsection{Step 1: Alignment}
To start with, let us focus on the alignment step. The goal of our multi-agent alignment is to compute a permutation matrix $\mathbf{P}()$ for each example which minimizes the distance between this example and the template of this state. 

In each state, the template $\mathbf{X}_{\mathrm{template}}$ contains a canonical spatial ordering of agents. Our learning method utilizes a general EM approach, which learns the template in an unsupervised way. The template learning process in a given class is shown in Algorithm \ref{alg:template_learn}.

Once the template is obtained, the permutation matrix for a given example in this state can be computed:
\begin{equation}
\argmin_{\mathbf{P}} ||\mathbf{P}(\mathbf{X})- \mathbf{X}_{\mathrm{template}}||_2
\end{equation}

This is a linear assignment problem and can be solved by computing a cost matrix for agents matching. Given the raw tracking data of $M$ agents, let us denote $\mathbf{D}$ as a $M\times M$ cost matrix that contains the Euclidean distance between each agent-agent location pair. $\mathbf{D}(m, n)$ indicates the distance between agent $m$ in the template and the agent $n$ in the given example, which is the cost of this agent-agent assignment.
\begin{equation}
\mathbf{D}(m,n)=||\mathbf{X}_{\mathrm{template}}(m)-\mathbf{X}(n)||_2
\end{equation}
Once the cost matrix $\mathbf{D}$ is computed, the Hungarian algorithm \cite{kuhn1955hungarian} is used to find the permutation matrix $\mathbf{P}()$ that minimizes the overall assignment cost. 

\begin{figure}[t]
\centering
\includegraphics[width=0.8\linewidth]{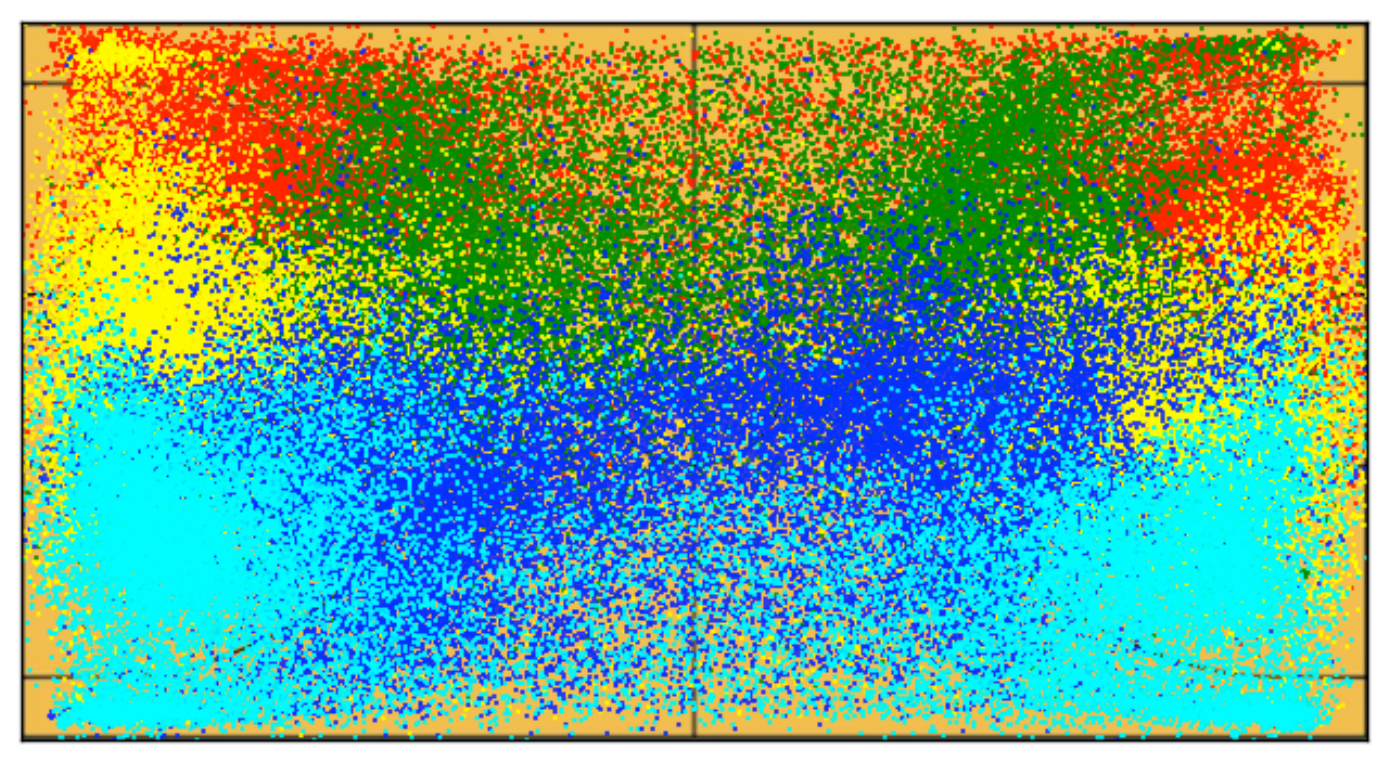}
\caption{The player distribution after our tree-based alignment. Our method reveals more detailed formation at each side of the court.}
\label{fig:tree_align}
\end{figure}


\subsection{Step 2: Data Partitioning}

Since the alignment has been solved, partitioning the data (i.e., clustering) into distinct states can be performed: 
\begin{equation}
\argmin_{\mathcal{C}}\sum_{C_k\in\mathcal{C}}\sum_{\mathbf{X}_i,\mathbf{X}_j \in C_k} ||\mathbf{P}(\mathbf{X}_i)-\mathbf{P}(\mathbf{X}_j)||_2\\
\label{eq:cluster_obj}
\end{equation}
where $\mathcal{C}$ is a set of clusters and $C_k$ represents the $k$-th data cluster. The clustering operation splits the data into more specific states and enable finer alignment to occur. 

As a clustering problem, we need to define the number of clusters to use. To determine this number of clusters, the specific application needs to be considered. For our case, we are interested in play retrieval, which means we need to balance two things: i) the total number of clusters is small, but still retaining high within-cluster similarity, and ii) the number of plays in each cluster high enough so we have enough plays to retrieve but small enough so that responsive retrieval can occur. To aid with this balancing act, we use an additional term which is similar to the idea of Silhouette analysis \cite{rousseeuw1987silhouettes} to constrain the number of clusters in each node of our tree:
\begin{equation}
E(\mathcal{X})=\frac{1}{|\mathcal{X}|}\sum_{C_{k} \in \mathcal{C}} \sum_{\mathbf{X}_i \in C_{k}} \frac{||\mathbf{P}(\mathbf{X}_i)-\mu_{kn}||_2-||\mathbf{P}(\mathbf{X}_i)-\mu_{k}||_2}{||\mathbf{P}(\mathbf{X}_i)-\mu_{kn}||_2}
\label{eq:constrain}
\end{equation}
where $\mu_{k}$ represent the mean of the cluster that example $\mathbf{X}_i$ belongs to and $\mu_{kn}$ indicates the mean of the closest neighbor cluster of example $\mathbf{X}_i$. Equation \ref{eq:constrain} measures the dissimilarity between neighboring clusters and how tightly the data is grouped within each cluster. When the number of clusters becomes too large, the similarity between neighboring clusters increases and $E$ decrease as well. Thus, we want to maximize $E$ to have the most discriminative clusters. 

For the data partitioning in each node, we attempt $K$-means clustering with $K$ equals to 2 to 10 and for each $K$ we compute the score $E$. The $K$ that provides the maximal $E$ will be selected to split the data in the current node.

\subsection{Tree Growth}

Since we can both align and cluster the multi-agent data, we can now learn the tree.  To clarify the notation, we use the subscript and superscript to indicate the node index and layer index. $\mathcal{X}_n^l$ indicates the data group in the $n$th node of the $l$th layer, $\mathcal{C}_n^l$ indicates the classes found from that node $\mathbf{X}_{\mathrm{template},n}^l$ represents the template in that node and $\mathbf{P}_n^l$ is the permutation matrix computed by using that template. For every node, we first use the Algorithm \ref{alg:template_learn} to align the data that is assigned to this node and then apply the clustering technique to split them into finer states. 

It is worth noting that the templates in each node should also be aligned so that the consistency of agents permutation can be preserved. Thus, the new template of node $n$ at layer $l$ is aligned to its parent template in the previous layer $\mathbf{X}_{\mathrm{template},n}^{l}=\mathbf{P}_{np}^{l-1}(\mathbf{X}_{\mathrm{template},n}^{l})$. it aligns the current template $\mathbf{X}_{\mathrm{template},n}^l$ to its parent template $\mathbf{X}_{\mathrm{template},np}^{l-1}$.  Then the same process repeat for each node in our tree. Algorithm \ref{alg:tree} summarizes the learning process of our tree-based method. During the learning process, the clusters $\mathcal{C}$ and templates $\mathcal{T}$ at each layer are stored for aligning process.   

\begin{algorithm}[t]
\caption{Learning process of tree-based alignment}\label{alg:tree}
\begin{algorithmic}[1]
\Procedure{LearnTree($\mathcal{X}$)}{}
\State $\mathcal{T}=\emptyset, \mathcal{C}=\emptyset$
\For{each layer $l$} 
\For{each node $n$}
\State learn template $\mathbf{X}_{\mathrm{template},n}^l$  using Algorithm \ref{alg:template_learn}
\State align to parent $\mathbf{X}_{\mathrm{template},n}^l=\mathbf{P}_{np}^{l-1}(\mathbf{X}_{\mathrm{template},n}^{l})$
\State align data $\mathbf{X}_i=\mathbf{P}_{n}^{l}(\mathbf{X}_i)$  $\forall \mathbf{X}_i \in \mathcal{X}_n^l$
\EndFor
\State store $[\mathbf{X}_{\mathrm{template},1}^l,...,\mathbf{X}_{\mathrm{template},N}^l]$ in $\mathcal{T}$
\State compute reconstruct loss with Eq. \ref{eq:cluster_obj}
\State terminate when stop criterion satisfies
\For{each node $n$}
\State Conduct K-means on $\mathcal{X}_n^l$ with different $K$
\State Select cluster set $\mathcal{C}_{n}^l$ that maximizes $E$ 
\State partition $\mathcal{X}_n^l$ to child nodes according to $\mathcal{C}_{n}^l$
\EndFor
\State Store $[\mathcal{C}_{1}^l,...,\mathcal{C}_{N}^l]$ in $\mathcal{C}$
\EndFor\\
\Return {$\mathcal{T}$, $\mathcal{C}$}
\EndProcedure
\end{algorithmic}
\end{algorithm}

There are two stop criterion: 1) a pre-defined maximum number of examples in each leaf node, 2) a pre-defined depth. From Equation \ref{eq:cluster_obj}, we know that the reconstruction cost would reach minimum if we had infinite states $|\mathcal{C}|=\infty$. Thus, we aim to find a minimum depth that can provide a considerably low cost. We plot the overall cost of Equation \ref{eq:cluster_obj} at each layer in Figure \ref{fig:depth_test} and set the minimum depth to 6 layers as it shows as the elbow point. A much deeper tree may be built for fast retrieval, but a 6-layer tree can achieve reasonable performance for alignment purpose. Figure \ref{fig:leaf_heatmap} shows an example of one leaf node. The top left image shows the centroid of the leaf node and others shows the distribution heat map of ball and each agent.

In terms of applying the tree-based alignment, given an input play, the player permutation is aligned to the global template at the root node first, it then moves to a child node by finding the nearest neighbor and repeats the alignment again. The aligned data in our tree-based method can be expressed as:
\begin{equation}
\mathbf{X}_{\mathrm{aligned}}~=\mathbf{P}^L(...(\mathbf{P}^2(\mathbf{P}^1( \mathbf{X_{\mathrm{raw}}}))))
\end{equation}
where $\mathbf{P}^l$ represents the permutation matrix at layer $l$. Essentially, such composition of permutation matrices yields the optimal ordering of the multiple agents. Figure \ref{fig:tree_align} shows the location distribution of each aligned agent across a quarter. By contrast to Figure \ref{fig:alignment}(c), our method reveals more meaningful structure at each side of the court.

 \begin{figure}[t]
\centering
\includegraphics[width=\linewidth]{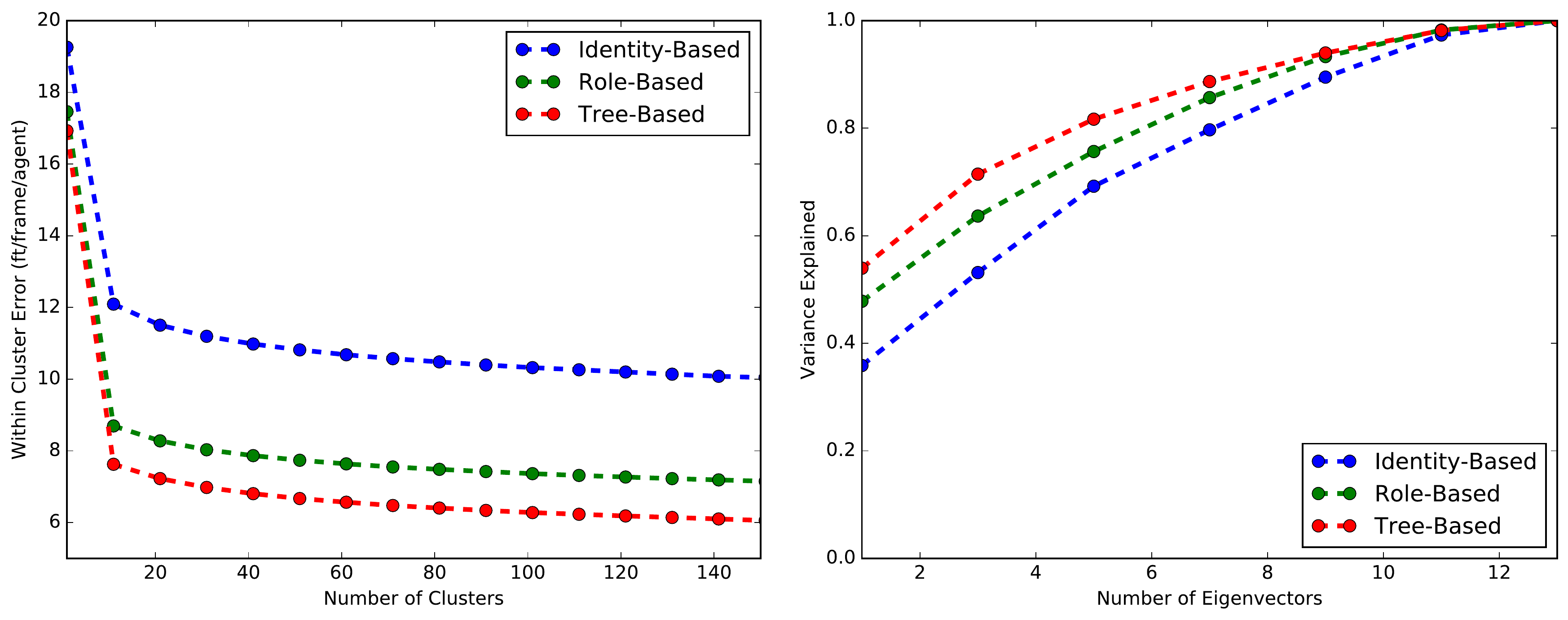}
\caption{The compressibility test results of using our tree-based alignment, role-based alignment and naive identity-based alignment. It shows our method provide the best compressibility.}
\label{fig:compress}
\end{figure}

\subsection{Alignment Evaluation}

Since better alignment should result in a more compressed input feature, we can evaluate our tree-based alignment via clustering and principle component analysis (PCA). To make clustering a fair comparison, instead of using the clusters generated by our approach inherently, K-means clustering is applied to both alignment methods. Given 100,000 frames, they are aligned with the role-based method and our method separately. Then, we apply K-means clustering to inspect the average within-cluster-error (WCE) with different K's 

\begin{equation}
\text{WCE}=\frac{1}{|\mathcal{X}|}\sum_{C_k}\sum_{\mathbf{X}_i \in C_k} ||\mathbf{X}_i-\mu_k||_2
\end{equation}
here we abuse $C_k$ to indicate the $k$th cluster after K-means clustering. PCA is used to inspect the variance explained by different eigenvectors and we calculate the variance via
\begin{equation}
\text{Variance~Explained} = \frac{\lambda_k}{\sum_{i=1}^D \lambda_i}
\end{equation}
where $\lambda_i$ is the $i$th eigenvalue that indicates the significance of the $i$th eigenvector. 

Apart from the role-based alignment and our tree-based alignment, the naive identity-based alignment is also used as a baseline. Identity-based approach only compares the trajectories according to players' identities which refer to an initial logical ordering (i.e., the player most like the point-guard ordered first, then the shooting-guard to the center - with this ordering fixed). Figure \ref{fig:compress} (left) shows the result of our clustering test and on the right, shows the performance using PCA. Both results show that our alignment gives the better compressibility.

\section{Fine-Grained Retrieval System}

The prime motivation of obtaining better alignment of the raw multi-agent data is to achieve better fine-grained retrieval. As depicted back in Figure~\ref{fig:query_example}, instead of typing a textual description of the play, users can selects an (a) example or modified example (b). The initial idea was first proposed in~\cite{sha2016chalkboarding}, where they utilized a simple hash-table by only clustering the ball trajectories.  Although effective, that approach is not an optimal solution since such hashing ignores the information of players. In this section, we show that we obtain better fine-grain retrieval utilizing our tree-based approach. We call the approach in ~\cite{sha2016chalkboarding} as the \emph{Baseline Method}.

\begin{figure}[t]
\centering
\includegraphics[width=\linewidth]{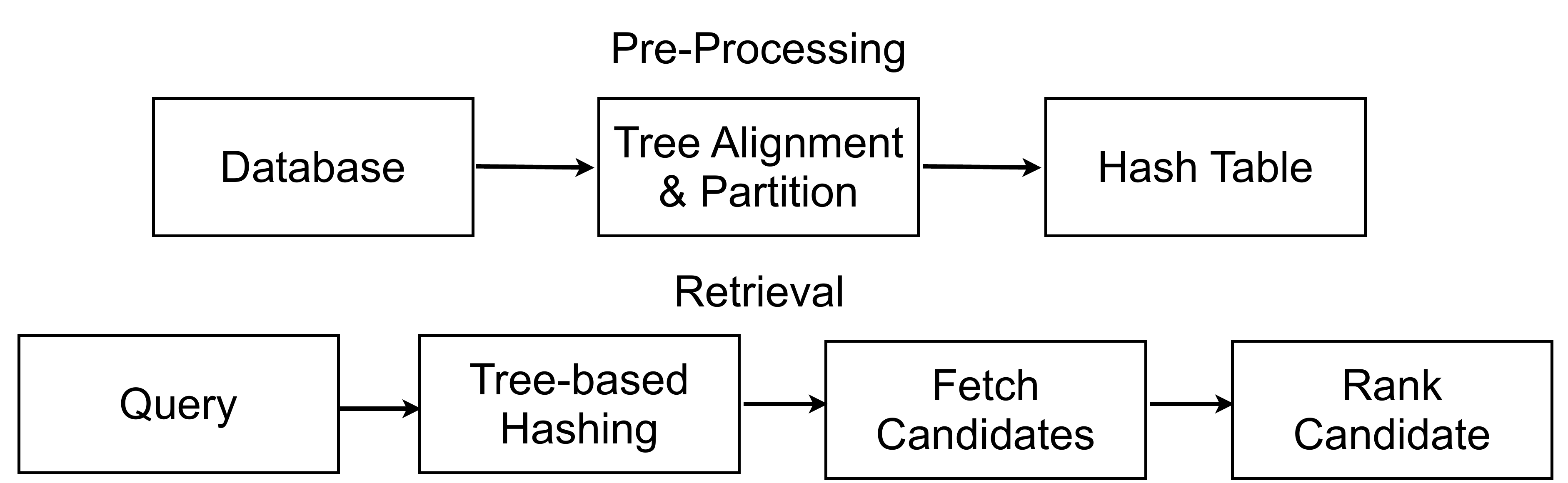}
\caption{The overview of our retrieval system. The top part shows the pre-processing and the bottom part shows the retrieval process.}
\label{fig:retrieval_flow}
\end{figure}

\begin{figure*}[t]
\centering
\includegraphics[width=\linewidth]{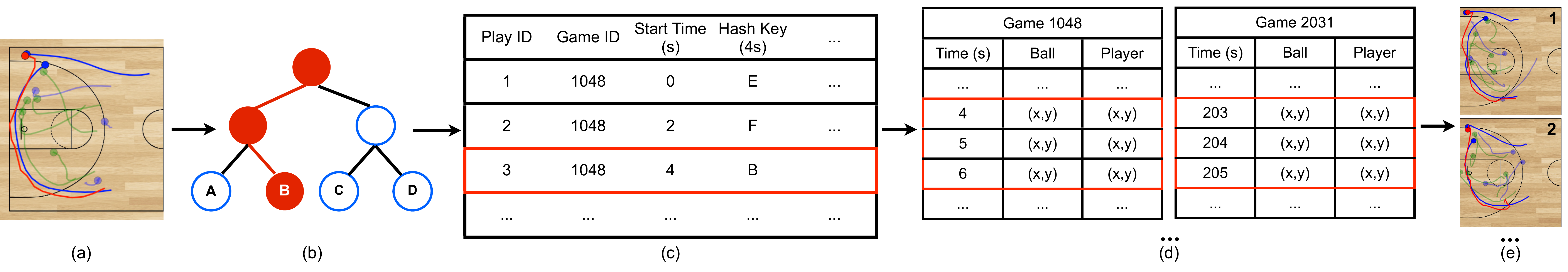}
\caption{The retrieval process of our system. (a) The input query. (b) Tree-based hashing which aligns the query and computes the hash key. (c) Find candidates with the hash key. (d) Fetch plays from the database. (e) Rank and return all the results.}
\label{fig:retrieval_process}
\end{figure*}

\begin{figure*}[t]
\centering
\includegraphics[width=0.9\linewidth]{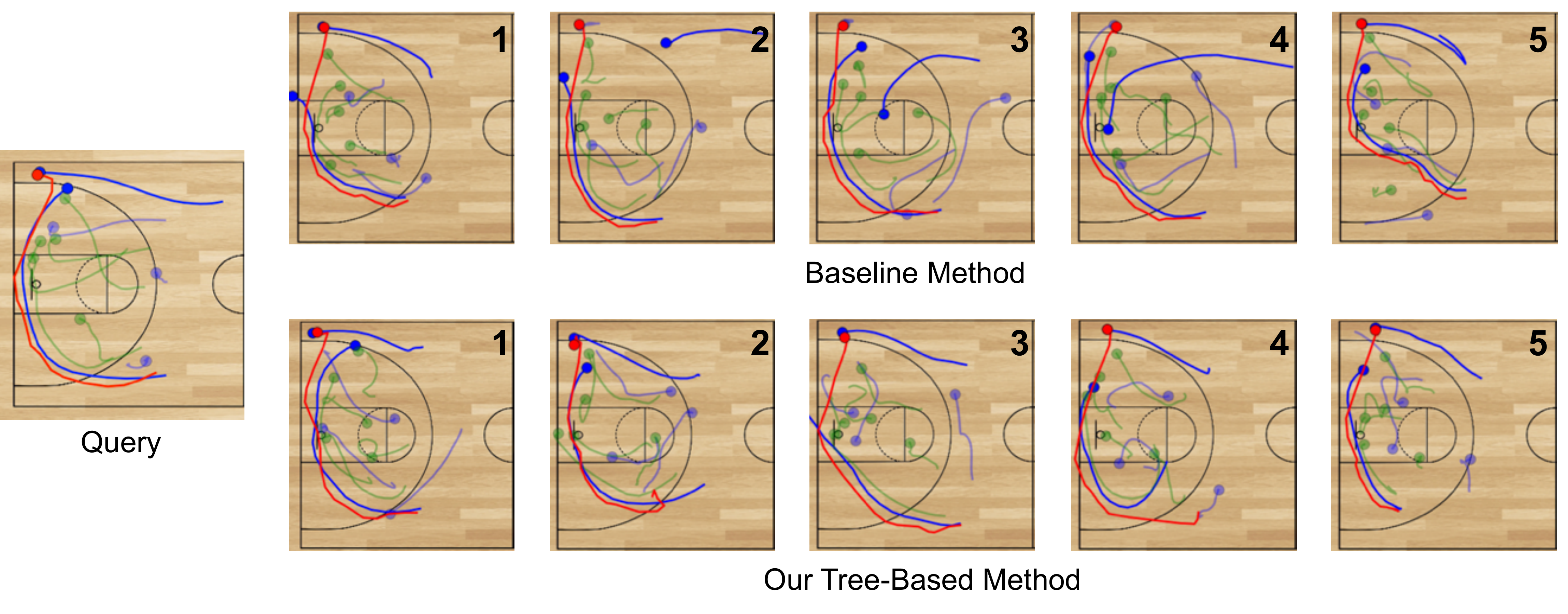}
\caption{A retrieval example of using two methods. The top row shows the top-5 results returned by the baseline method~\cite{sha2016chalkboarding} while the bottom row shows the top-5 results returned by our method.}
\label{fig:retrieval_example}
\end{figure*}

The dataset that is used in this work is the SportVU basketball dataset. The dataset is captured by the STATS SportVU system~\cite{stats2015sportvu}, which generates location data for every player and ball at 25Hz, along with detailed logs for actions such as passes, shots, fouls, etc. The dataset is taken from 1300 games from the last two seasons of a professional basketball league. 1200 games are used as our database and queries are extract from the rest 100 games. The tracking data of each game is stored in a separate table and each row contains the information of one player at one frame, which are time, team ID, player ID, action ID and the $(x,y,z)$ location of the player or the ball. Although our retrieval focuses on short plays, the database is still stored in its original form. The information of plays are saved in our hash table. 

A block digram of our retrieval system is given in Figure~\ref{fig:retrieval_flow}. In consists of two parts: i) pre-processing and ii) retrieval. In terms of pre-processing, we first extract all the plays with fixed lengths (i.e., 1, 2, 3, 4 and 5secs continuous chunks). For each input play $\mathbf{X}_{\mathrm{play}}$, we then find the permutation matrix according to our tree which finds the best ordering and then store that permutation matrix for that play. To enable fast play retrieval, we then associate each play to a hash-key which is found using our alignment/clustering method describe in the previous section. Both permutation matrices and hash keys are stored in our hash table. Our hashing method is similar to the concept of locality sensitive hashing (LSH) since similar plays are placed to the same address.

In terms of retrieval, given an input play we first align it to the tree of templates. We then return all the plays that are associated with the hash-key (we call these the \emph{candidate plays}), which are associated with the leaf-node of the template used to align the query at the lowest level. We then determine the similarity of the plays depending on which players were selected by the user for fine-grain interactive search and rank display the ranked results to the user. 

An example of the query process is shown in Figure~\ref{fig:retrieval_process}. Given a 4-second input query (which has the ball and the two offensive players selected) (a), we then put input query into our tree (b). Within the tree, we traverse the path through the tree until a leaf-node is reached (which is ``B'' as depicted by the red path). Based on that hash-key ``B'', all the plays with that hash-key are then retrieved from the hash table (c) and fetched from the database (d). Similarity between the aligned input query and the plays in the database with that hash-key are then computed - with only the trajectories selected contributing to the similarity score. Once a similarity score has been calculated, a ranked list of the plays in order of similarity (or smallest $L_{2}$ distance between the trajectories) is given (e). Some additional fixed weightings depending on the team-ID or recent plays can also be included once that initial ranked list has been generated.   


Figure \ref{fig:retrieval_example} displays the top-5 results for the baseline method and our tree-based method given the input query. This examples qualitatively highlights the benefit of our approach as it shows that the baseline method can not find the corresponding players correctly due to the imperfect alignment while our method maintains a high consistency between results and the query. In the next section, we show the results of an user study which quantitatively shows that the tree-based method yields better interactive fine-grain retrieval.

\section{User Experiments}

\subsection{Experiment Design}

To show the benefit of our our tree-based alignment method to the previous method described in~\cite{sha2016chalkboarding} -- which we call the \emph{Baseline Method} -- we conducted a series of user studies which focused on the task of interactive fine-grain retrieval (i.e., where the user selects a subset of players within an example play). To enable a fair comparison, whilst also maintaining responsive retrieval times ($<1sec$), we set the maximum size of plays within a leaf node to 2000, which generates a deeper tree with 314 individual leaf nodes/hash entries. Both the baseline method and the tree-based method utilized 314 clusters. 



\begin{figure}[t]
\centering
\includegraphics[width=\linewidth]{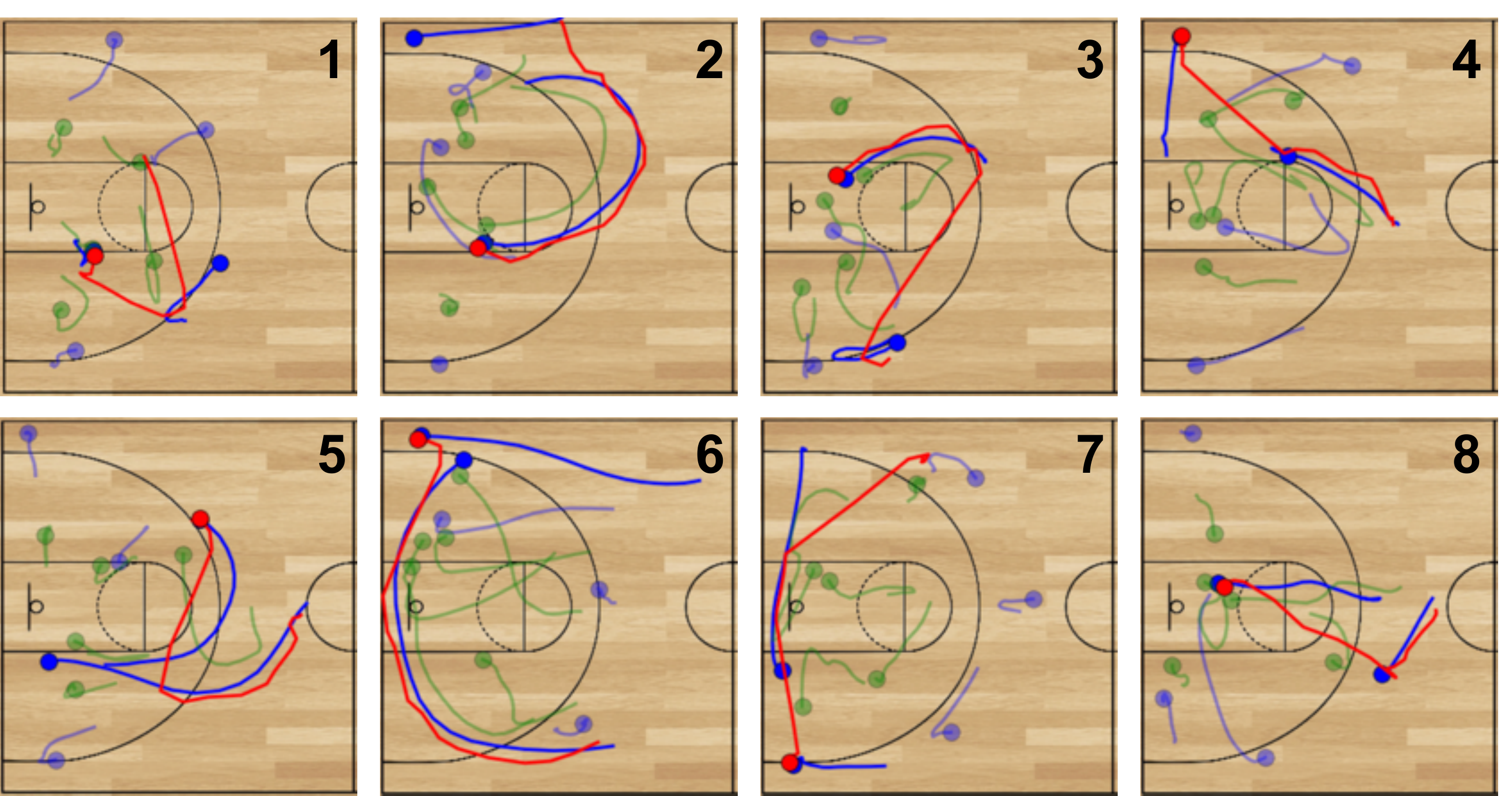}
\caption{Eight queries that are used in our user study, where blue is the offense team, green is the defense team and red represents the ball. The highlighted trajectories shows the selected players for the third retrieval setting.}

\label{fig:experiment_queries}
\end{figure}

Eight retrieval tasks were selected for our user study and we tested three different settings on each tasks.
\begin{enumerate}
\item Retrieval conditioned on all the players and the ball
\item Retrieval conditioned on the offense team and the ball
\item Retrieval conditioned on two selected players and the ball
\end{enumerate}
Because the baseline method requires the ball trajectory for its hashing function, ball is included in each setting. An user study is conducted for each setting and Figure~\ref{fig:experiment_queries} shows the eight queries for the third setting. 

\begin{figure}[t]
\centering
\includegraphics[width=0.8\linewidth]{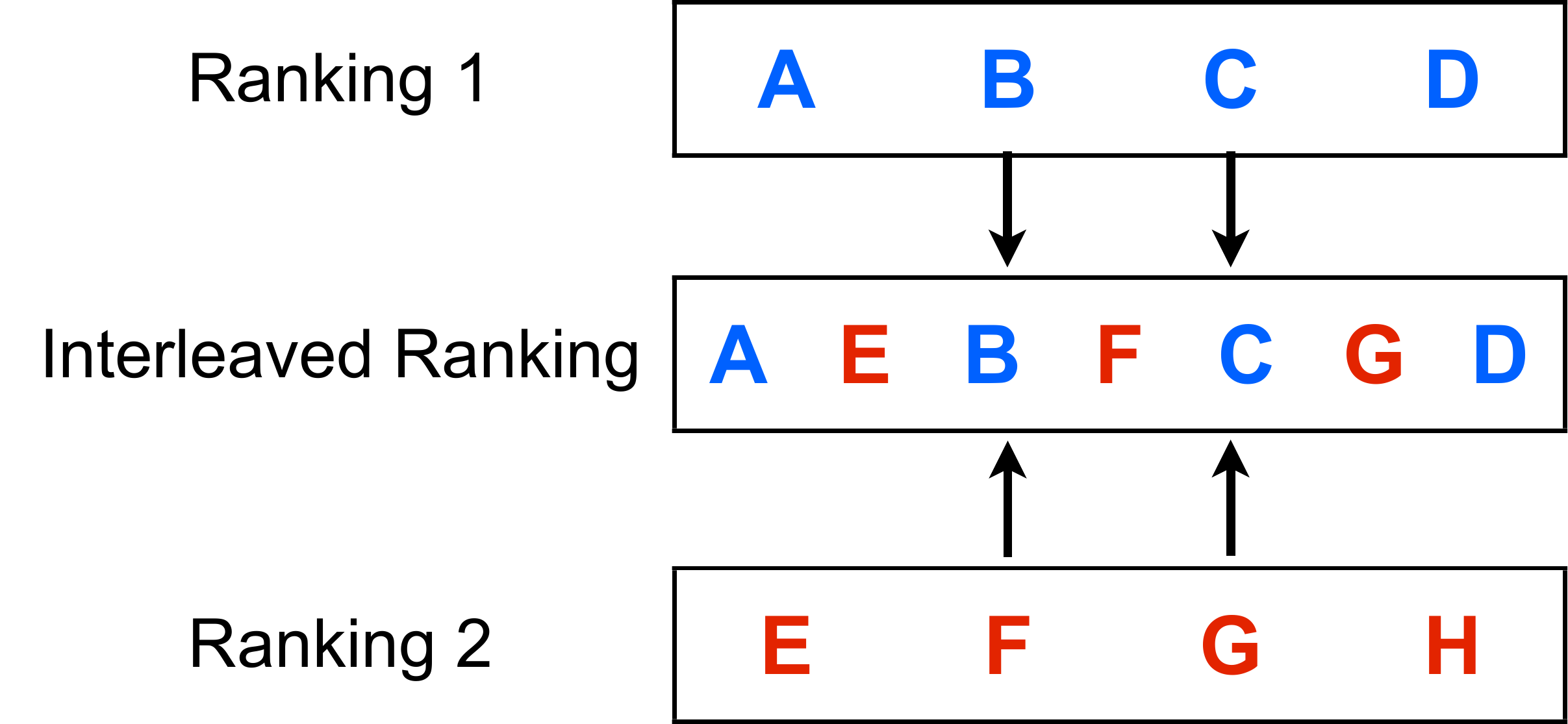}
\caption{Depicting an interleaving of two rankings.}
\label{fig:interleave_rank}
\end{figure}

We evaluated the retrieval quality via an interleaved evaluation, where the top 10 results returned by the baseline method and our method were combined via the Team-Draft Interleaving method~\cite{chapelle2012large} into a single ranking (see Figure~\ref{fig:interleave_rank}). The combined ranking and its query were then displayed in an online survey form so that users could view the results top-down and select the relevant plays.

\subsection{Procedure}

We recruited ten people with strong basketball background to participant our user studies. For each study, every person spent their first 5 minutes reading the instruction, which helped them to understand the images in Figure~\ref{fig:experiment_queries} and how to select relevant plays. After that, half an hour is allocated for each participant to finish all eight questions. Such procedure was repeated for each retrieval setting.

Each question contains one input query and interleaved retrieved results from two systems. If one result was returned by both systems, it would only be displayed once in our survey. Participants were asked to scan the results top-down and check the plays that they think are relevant to the input query. 

\subsection{Benchmark Results}

Using the relevance feedbacks from the participants, we performed a benchmark comparison by using the average precision and the expected reciprocal rank of the first result, which are two standard retrieval evaluation metrics~\cite{chapelle2009expected,manningh,salton1986introduction}. Let $r_j$ denote the rank of the $j$-th relevant document, then the average precision of a ranking list can be computed by:
\begin{equation}
AvgPrec=\frac{1}{\#rel}\sum_j Prec@r_j
\end{equation}
where $Prec@r_j$ is the precision of the top $r_j$ items in the ranking list, and $\#rel$ is the number of relevant items in the ranking list. The expected reciprocal rank is simply the inverse of the rank of the first relevant result, which can be computed by:
\begin{equation}
ERR=\frac{1}{r_1}
\end{equation}
The expected reciprocal rank is more sensitive to the efficiency of finding the first result while average precision is more recall-focused. For our user study, we computed both of them on the two ranking lists embedded in the interleaved ranking, and over the pooling of both top-10 results. 

Table \ref{table:user_study_1} shows the result of average precision in three different retrieval settings. In each setting, the top two rows show the mean average precision aggregated across all ten users for each method. It shows that our method has higher precision than the baseline method in all three settings, and the improvement of our method becomes larger when the queries become more specific (fewer agents are involved). The ``Win/Loss'' rows at bottom indicate how many individual participants have higher average precision with our method. It shows our tree-based approach wins in most users.

\begin{figure}[t]
\centering
\includegraphics[width=\linewidth]{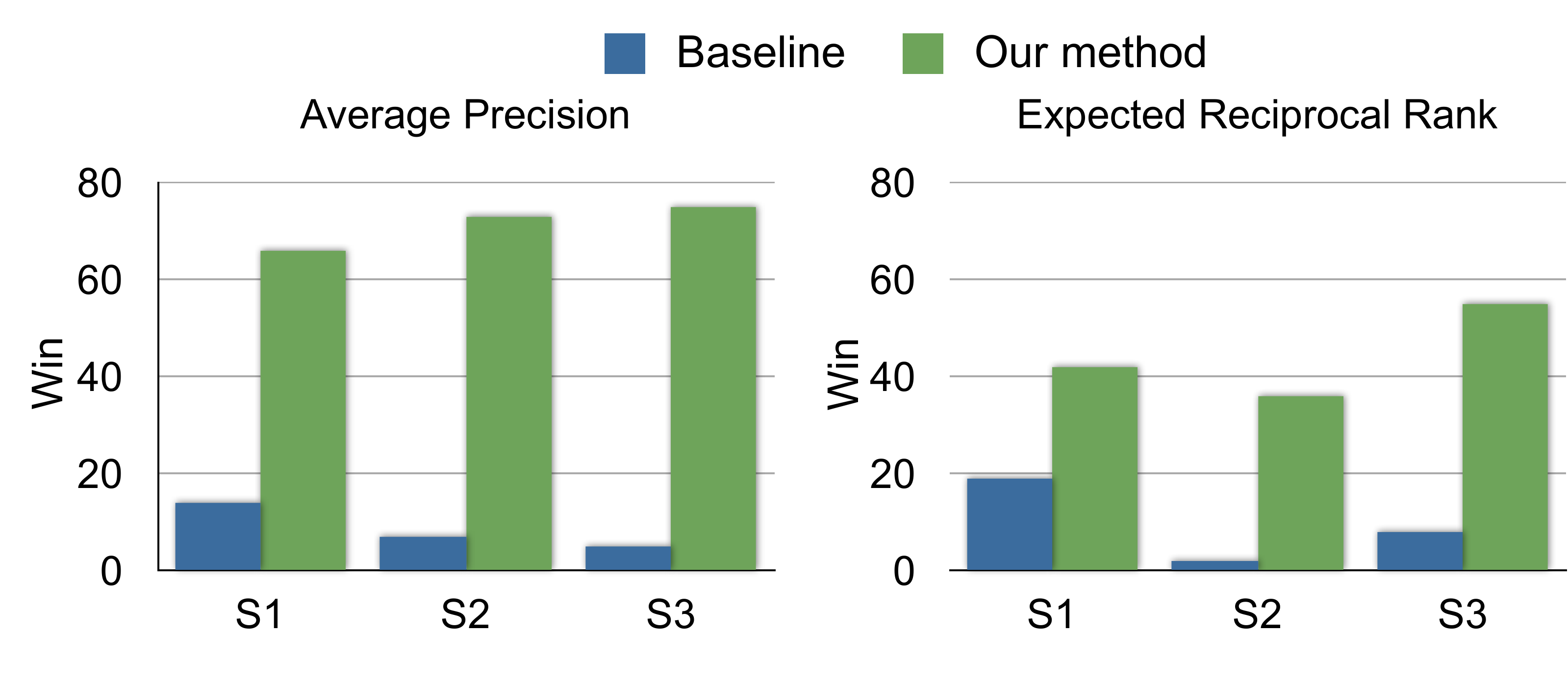}
\vspace{-15pt}
\caption{These bar charts indicate the overall win rate of our method with each metric and retrieval setting. They show that our method outperforms the baseline in all situations.}
\label{fig:bar_chart}
\end{figure}

Table \ref{table:user_study_2} compares expected reciprocal rank, which has the same structure to Table \ref{table:user_study_1}. Please note that since both methods may have the first relevant result at the same rank, a draw could occur sometimes. Thus, each cell in the ``Win/Loss'' rows may not sum to 10. Similarly, our method outperforms the baseline method especially in the third setting (S3). It shows that using one template cannot find the corresponding agents correctly when queries become very specific. The bar charts in Figure \ref{fig:bar_chart} highlight the overall ``Win/Loss'' in Table \ref{table:user_study_1} and \ref{table:user_study_2}.

\section{Related Work}

\begin{table}[t]
\caption{This table compares the mean average precision aggregated across all users for each query (Q1-Q8) with three different retrieval settings (S1-S3).  The ``Win / Lose'' rows show the number of users for whom the tree-based approach achieved a higher average precision. }{
\centering
\resizebox{\linewidth}{!}{
\begin{tabular}{l | l | c c c c c c c c | c}
\hline
\hline
& & Q1 & Q2 & Q3  & Q4  & Q5 & Q6 & Q7 & Q8  & Overall \\
\hline  \multirow{3}{*}{S1} & Baseline Method  & 0.12 & 0.07 & 0.08 & 0.25  &  0.15 & 0.25 & 0.24 & 0.37 & 0.19 \\
   \cline{2-11} & Tree-Based & 0.57 & 0.67 & 0.50 & 0.34 & 0.72 & 0.58 & 0.39 & 0.42 & \textbf{0.52}\\
   \cline{2-11}
   & \textbf{Win / Lose} & 10 / 0 & 10 / 0 &  10 / 0 & 6 / 4 & 10 / 0 & 10 / 0 &  5 / 5 & 5 / 5 & \textbf{66 / 14} \\
\hline  \multirow{3}{*}{S2} & Baseline Method  & 0.12 & 0.09 & 0.08 & 0.18  &  0.27 & 0.36 & 0.21 & 0.37 & 0.21 \\
   \cline{2-11} & Tree-Based & 0.77 & 0.70 & 0.81 & 0.57 & 0.73 & 0.58 & 0.50 & 0.47 & \textbf{0.65}\\
   \cline{2-11}
   & \textbf{Win / Lose} & 10 / 0 & 10 / 0 &  10 / 0 & 10 / 0 & 8 / 2 & 8 / 2 &  9 / 1 & 8 / 2 & \textbf{73 / 7} \\
   \hline  \multirow{3}{*}{S3} & Baseline Method  & 0.04 & 0.07 & 0.12 & 0.08  &  0.23 & 0.09 & 0.21 & 0.15 & 0.12 \\
   \cline{2-11} & Tree-Based & 0.87 & 0.78 & 0.68 & 0.71 & 0.70 & 0.73 & 0.56 & 0.58 & \textbf{0.70}\\
   \cline{2-11}
   & \textbf{Win / Lose} & 10 / 0 & 10 / 0 &  10 / 0 & 10 / 0 & 9 / 1 & 10 / 0 &  6 /4 & 10 / 0 & \textbf{75 / 5} \\
\end{tabular}}}
\label{table:user_study_1}
\end{table}
\vspace{-5pt}

\begin{table}[t]
\caption{This table compares the expected reciprocal rank aggregated across all users for each query (Q1-Q8) with three different settings (S1-S3).  The ``Win / Lose'' rows show the number of users for whom the tree-based approach found a relevant result earlier in the ranking. }{
\centering
\resizebox{\linewidth}{!}{
\begin{tabular}{l | l | c c c c c c c c | c}
\hline
\hline
& & Q1 & Q2 & Q3  & Q4  & Q5 & Q6 & Q7 & Q8  & Overall \\
\hline  \multirow{3}{*}{S1} & Baseline Method  & 0.5 & 0.44  & 0.33 & 0.58  &  0.43 & 0.82 & 0.66 & 0.75 & 0.56 \\
   \cline{2-11} & Tree-Based & 0.75 & 0.94 & 0.55 & 0.49 & 0.94 & 1 & 0.73 & 0.91 & \textbf{0.79}\\
   \cline{2-11}
   & \textbf{Win / Lose} & 5 / 2 & 9 / 0 & 8 / 2 & 3 / 5 & 7 / 0 & 2 / 0 &  5 / 3 & 3 / 1 & \textbf{42 / 19} \\
\hline  \multirow{3}{*}{S2} & Baseline Method  & 0.51 & 0.27 & 0.45 & 0.82  &  0.86 & 0.9 & 0.71 & 0.80 & 0.66 \\
   \cline{2-11} & Tree-Based & 1 & 0.83 & 1 & 0.86 & 1 & 0.9 & 0.85 & 0.84 & \textbf{0.91} \\
   \cline{2-11}
   & \textbf{Win / Lose} & 5 / 0 & 8 / 0 &  10 / 0 & 2 / 0 & 2 / 0 & 2 / 0 &  4 / 1 & 3 / 1 & \textbf{36 / 2} \\
   \hline  \multirow{3}{*}{S3} & Baseline Method  & 0.2 & 0.36 & 0.48 & 0.37  &  0.63 & 0.32 & 0.75 & 0.54 & 0.46 \\
   \cline{2-11} & Tree-Based & 1 & 1 & 0.76 & 1 & 0.95 & 1 & 0.71 & 0.9 & \textbf{0.92} \\
   \cline{2-11}
   & \textbf{Win / Lose} & 10 / 0 & 8 / 0 &  6 / 1 & 9 / 0 & 7 / 1 & 9 / 0 &  0 / 4 & 6 / 2 & \textbf{55 / 8} \\
\end{tabular}}}
\label{table:user_study_2}
\vspace{-5pt}
\end{table}

Information retrieval has a long research history in computer science domain~\cite{schutze2008introduction,chowdhury2010introduction} and the majority of those previous studies focus on  tokenized query format. Although tokenized query has been widely used to both text data and multimedia data~\cite{kim2013time,xia2013online,zhuang2011two}, some research has shown that using free-form or ``ad-hoc" queries can be significantly more user-friendly~\cite{manningh,salton1986introduction}. One popular free-form query type in recent research is the exemplar-based/sketch-based query format~\cite{bui2015scalable,yu2016sketch,zhang2016sketchnet} in image retrieval. Such query format enables users to issue the query at a more intuitive and precise level. 

In terms of sports analytics domain, most works focus on evaluating and comparing players performance~\cite{franks2015counterpoints, chen2016modeling}, analyzing broadcasting videos~\cite{liu2005effective} and discovering behavior patterns/styles~\cite{miller2014factorized,yue2014learning,wei2016forecasting}. Similar to other domains, the conventional approach of sports data retrieval still uses directory/taxonomy paradigm~\cite{wall2011history} to categorize sports plays~\cite{mcqueen2014automatically,chen2014play,bialkowski2014win}. Since multi-agent spatiotemporal data has been widely collected, using the sketch-based or exemplar-based query can be a more effective and user-friendly solution. The first formal spatiotemporal query paradigm is purposed by Sha et al.~\cite{sha2016chalkboarding}. They developed a new interface that accepts exemplar-based or sketch-based queries for sports play retrieval. It was reported that their query format is much more effective and user-friendly than the conventional text-based retrieval.

From the technical perspective, the primary challenge of retrieving  multi-agent spatiotemporal data is how to compare them effectively. Although similarity measure has been well investigated on trajectories and time series ~\cite{chen2007spade,toohey2015trajectory,eichmann2015evaluating}, most of them only focused on single trajectories rather than multi-agent ones. The seminal work of comparing multi-agent data called ``role-based" representation~\cite{lucey2013representing, wei2013large}. It uses a formation template to order the agents so that corresponding agents can be found between two samples. However, this method is suboptimal because using only one template is agnostic to those fine-grained behaviors. 

Apart from the permutation alignment of multi-agent data, we still need to find the an effective similarity measure between individual pairs of trajectories. There are two main categories in trajectory comparison studies, one of them focuses on elastic measure that addresses shifting and warping issues in both time and space domains~\cite{listgarten2005multiple,chen2007spade,keogh2005exact,keogh2000scaling}, while the other group of research focuses on finding the most similar or dissimilar points between two trajectories, which ensures the robustness~\cite{lou2002semantic,junejo2004multi}. Euclidean distance is used in our work because the experimental result in ~\cite{sha2016chalkboarding} shows that Euclidean distance is still the most effective metric for trajectory comparison in sports. 

In all modern retrieval system, fast indexing is required for fast search through a large database. Hash table is one of the most popular approach to achieve this purpose~\cite{blanco2015fast,zhang2015query,liu2017semantic}. Hash function is normally designed for specific domain and application, but in general, it aims to reduce the time cost. Similar to~\cite{sha2016chalkboarding}, our method uses the concept of locality sensitive hashing (LSH)~\cite{indyk1998approximate,gionis1999similarity}, which is designed to place similar samples into a same address. Such method has been applied in other settings where similarity measure or ranking is required~\cite{setty2017modeling,liu2017semantic}.

\vspace{-3pt}
\section{Conclusion, Discussion and Future Work}
In this paper, we presented a new tree-based alignment method which enables effective similarity measure between multi-agent trajectories data. Based on this method, we also presented a retrieval system tailored towards accurate and efficient sports play retrieval. A compressibility experiment showed that our tree-based method outperforms the state-of-the-art alignment method, and our full-stack retrieval system demonstrate its effectiveness in an user study where our approach achieve a higher precision than the baseline method. From a relevance estimation standpoint, even though our alignment has improved the similarity measure, the choice of distance metric can be improved. For instance, the concept of ``inverse document frequency''~\cite{salton1986introduction} in information retrieval indicates that tokens that appear more frequent in document are not as indicative of relevance as those more rare words. Thus, similar idea can be incorporated into trajectory distance measure. More generally, machine learning techniques can be applied to learn a better distance metric with appropriate training data. Apart from retrieval, our tree-based alignment could be applied to other important data mining tasks. Since a playbook is learned inherently by our tree, it can be used in game summarization and team/player characterization where those game states are required. Beyond sports, our method could be applied to a wide range of domains where multi-agent spatiotemporal data is involved. One example could be the crowd behavior analysis in surveillance domain.
\bibliographystyle{ACM-Reference-Format}
\bibliography{acmart} 

\end{document}